\documentclass[useAMS,usenatbib,pdfsync]{mn2e}

\usepackage{graphicx}

\newcommand       \be           {\begin{equation}}
\newcommand       \ee           {\end{equation}}
\newcommand       \bea          {\begin{eqnarray}}
\newcommand       \eea          {\end{eqnarray}}
\newcommand       \apj          {ApJ}

\newcommand       \aap          {A\&A}

\newcommand       \mnras        {MNRAS}
\def\simlt{\mathrel{\hbox{\rlap{\hbox{\lower4pt\hbox{$\sim$}}}\hbox{$<$}}}}
\def\simgt{\mathrel{\hbox{\rlap{\hbox{\lower4pt\hbox{$\sim$}}}\hbox{$>$}}}}

\title[Magnetar Winds and GRBs]{Relativistic Jets and Long-Duration
Gamma-ray Bursts from the Birth of Magnetars} \author[N. Bucciantini,
E. Quataert, J. Arons, B.~D. Metzger,
T.~A. Thompson]{N. Bucciantini$^{1}$\thanks{E-mail:
nbucciantini@astro.berkeley.edu}, E. Quataert$^{1}$,
J. Arons$^{1,2,3}$, B.~D. Metzger$^{1,2}$, T.~A. Thompson$^{4}$\\
$^{1}$Astronomy Department and Theoretical Astrophysics Center,
University of California, Berkeley, 601 Campbell Hall, Berkeley CA,
94720\\ $^{2}$Department of Physics, University of California,
Berkeley, Le Conte Hall, Berkeley, CA 94720 \\ $^{3}$Kavli Institute
for Particle Astrophysics and Cosmology, Stanford University\\
$^{4}$Lyman Spitzer Jr.~Fellow, Department of Astrophysical Sciences,
Peyton Hall, Ivy Lane, Princeton, NJ, 08544}
\begin{document}

\date{Accepted . Received ; in original form }

\pagerange{\pageref{firstpage}--\pageref{lastpage}} \pubyear{????}

\maketitle

\label{firstpage}

\begin{abstract}

We present time-dependent axisymmetric magnetohydrodynamic simulations
of the interaction of a relativistic magnetized wind produced by a
proto-magnetar with a surrounding stellar envelope, in the first $\sim
10$ seconds after core collapse.  We inject a super-magnetosonic wind
with $\dot E = 10^{51}$ ergs s$^{-1}$ into a cavity created by an
outgoing supernova shock.  A strong toroidal magnetic field builds up
in the bubble of plasma and magnetic field that is at first inertially
confined by the progenitor star.  This drives a jet out along the
polar axis of the star, even though the star and the magnetar wind are
each spherically symmetric. The jet has the properties needed to
produce a long-duration gamma-ray burst (GRB).  At $\sim 5$ s after
core bounce, the jet has escaped the host star and the Lorentz
factor of the material in the jet at large radii $\sim 10^{11}$ cm is
similar to that in the magnetar wind near the source.  Most of the
spindown power of the central magnetar escapes via the relativistic
jet.  There are fluctuations in the Lorentz factor and energy flux in
the jet on $\sim 0.01-0.1$ second timescale.  These
may contribute to variability in GRB emission (e.g., via internal
shocks).

\end{abstract}


\section{Introduction}
\label{sec:int}

Observations of long-duration gamma-ray bursts (GRBs) show that many
are associated with core-collapse supernovae (SNe) (\citealt{woo06}).
A newly formed rapidly rotating magnetar (e.g.,
\citealt{usov92,thom94,wheeler00}) or a black hole with an accretion
disk (e.g., \citealt{mcf99}) are possible central engines powering
GRBs and their associated SNe.  In both of these models, a key problem
is to understand how the relativistic material necessary for
generating the GRB escapes from deep within the host star.  A natural
possibility, suggested by both afterglow observations (``jet breaks'';
\citealt{rho99}) and GRB energetics, is that a collimated outflow
punches through the stellar envelope, providing a channel out of which
the relativistic material can escape (e.g., \citealt{mat03}).

In the collapsar model, collimated outflows from GRBs are accounted
for by jets produced by an accretion flow onto a central black hole.
In the magnetar model, the origin of such collimated outflows is less
clear because relativistic outflows by themselves do not efficiently
self-collimate (e.g., \citealt{le01}). \citet{kg02} suggested, by
analogy to pulsar wind nebulae \citep{beg92} (PWNe), that the
interaction of the wind from the spinning-down magnetar with the
surrounding star could facilitate collimation.  In a previous paper,
we used axisymmetric thin shell calculations to demonstrate that this
can in fact occur (\citealt{b08}; see \citealt{um06,um07} for related
ideas based on force-free rather than inertially loaded outflows).  As
K\"onigl and Granot anticipated, the physical picture is analogous to
that used to understand the morphologies of pulsar wind nebulae
(\citealt{kom04,ldz04}): the magnetar wind shocks on the surrounding
(exploding) stellar envelope, creating a bubble of relativistic plasma
and magnetic field inside the host star (a ``magnetar wind nebula'';
MWN).  If the toroidal magnetic field in the bubble is sufficiently
strong, the bubble expands primarily in the polar direction due to the
``tube of toothpaste'' effect - the magnetic hoop stress in the
equator creates a pressure gradient along the axis which is not
balanced by any restoring force, thus driving the flow preferentially
in the polar direction. The nebula itself is ultimately confined by
the inertia of the SN envelope, to which little energy is transferred,
in contrast to the pressure confinement in traditional magnetic tower models
(\citealt{she07}) or the case of confinement by a pressurized 
cocoon inside the progenitor star (\citealt{um06}).

Our previous results were based on a thin-shell model for the
interaction between a magnetar wind and its host star.  In this {\it
Letter}, we build on this work by carrying out time-dependent
axisymmetric magnetohydrodynamic (MHD) simulations of the interaction
between a magnetar wind and a surrounding star.  This paper is
organized as a proof of principle to show that a mechanism developed
in the PWNe context to explain the origin of observed jet/plume
features, can also apply to proto-magnetars.  For this reason we have
selected a simplified set of parameters, deferring a detailed
parameter study to future work.

We assume that an outgoing SN shock has already created a central
evacuated cavity and that the host star is spherically symmetric.  A
magnetar is assumed to inject energy at a constant rate of $\dot E =
10^{51}$ ergs s$^{-1}$, a spindown power that is appropriate for
proto-magnetars with $P \sim 1$ ms and $B \sim 10^{15}$ G (e.g.,
\citealt{met07}).  The interaction between the magnetar wind and the
host star depends on the strength of the toroidal magnetic field $B$
in the MWN; this in turn depends on the magnetization in the wind at
large radii, $\sigma=r^2B^2c/\dot{E}$. More specifically, 
$\sigma (R_{TS})=\sigma_{TS}$, the magnetization at radii typical of 
the termination shock's radius $R_{TS}$, is the essential control 
parameter. At early times after
core-bounce ($\simlt$ a few sec), young magnetars lose significant
amounts of mass to a neutrino driven wind, leading to $\sigma_{LC}
\simlt 1$, where $\sigma_{LC}$ is the magnetization evaluated at the
light cylinder.  However, the mass loss rate decreases as the neutron
star cools and the wind at later times ($\sim 10-100$ sec) is
ultra-relativistic with $\sigma_{LC} \sim 10^2-10^3$
(\citealt{thom04}; \citealt{met07}).  This late-time outflow is the
most promising for producing a GRB.  The relation between
$\sigma_{LC}$ -- which can be calculated with some confidence -- and
$\sigma$ at larger radii in the wind is difficult to determine because
of uncertainties in the conversion of magnetic energy into kinetic
energy in relativistic winds (see \citealt{b08} for a detailed
discussion of this problem in the current context).  In this {\it
Letter} we focus on later times when $\sigma_{LC}$ is large.  We
assume that non-ideal MHD processes analogous to those that operate in
the PSR-PWN problem are also at work in proto-magnetar winds; these
convert the high $\sigma_{LC}$ wind at the light-cylinder into a
moderate $\sigma_{TS}$,  high $\gamma$ wind at the free wind's termination
radius.  Specifically,
we consider relativistic winds with $\gamma_w = 10$ or $25$ and
$\sigma_{TS} = 0.1$; GRBs require somewhat higher $\gamma$ outflows but
these are more difficult to simulate for numerical reasons.  


The remainder of this paper is organized as follows.  In the next
section (\S \ref{sec:num}) we describe our numerical setup.  We then
present the results of our simulations (\S \ref{sec:res}) and discuss
their implications for models of GRBs (\S \ref{sec:con}).

\vspace{-0.9cm}

\section{Numerical setup}
\label{sec:num}

All of the simulations were performed using the shock-capturing
central-scheme for relativistic ideal MHD developed by \citet{ldz02}
\& \citet{ldz03}, using ideal gas EoS with adiabatic coefficint $4/3$.
 We refer the reader to these papers for a detailed
description of the equations and the numerical algorithms.

The interaction of the magnetar wind with the surrounding SN
progenitor is investigated by performing 2-D axisymmetric simulations
on a spherical grid. The domain in $\theta$ is the first quadrant from
$\theta = 0$ to $\theta = \pi/2$, with reflecting boundary conditions
for $v_\theta$ and $B$ at the polar axis to enforce axisymmetry, and
similar boundary conditions in the equatorial plane.  The grid is
uniform in the $\theta$ direction with 100 cells.  Given that one
needs to study a wide range of spatial scales, from the termination
shock at $\sim 10^8$ cm to the outer edge of the star at $\sim
2\,10^{10}$ cm, we chose a grid that is logarithmic in radius and that
extends from $r_{\min}=10^7$ cm to $r_{\max}=5 \, 10^{10}$ cm or
$r_{\max} = 10^{11}$ cm, with 100 cells per decade in radius.  Zeroth
order extrapolation is assumed at the outer boundary.  The code is
second order in both space and time, with a monotonized central
limiter, chosen in order to resolve the large density jump between the
lighter relativistic plasma inside the MWN and the heavier stellar
envelope (the density can increase by a factor of $\sim 10^4$).

For proto-magnetars with millisecond rotation periods, the location of
the Alfv\'enic surface is at $\sim 10^7$ cm and the fast magnetosonic
surface is at $\sim 10^7-10^8$ cm (\citealt{me06}; hereafter B06).  In
the present simulations, we do not attempt to resolve the dynamics of
the wind in the sub-magnetosonic region close to the neutron star.
Instead, at $r_{\min}$ we inject a super-magnetosonic wind with a
fixed Lorentz factor of $\gamma_w=10$ or $\gamma_w = 25$ and a
magnetization of $\sigma = 0.1$, a value appropriate to distances
on the order of the termination shock radius (the simulations were 
less stable for higher $\sigma$); under these assumptions, $\sigma$ is conserved thoughoutb the upstream region. The wind is assumed to be cold 
with $\rho c^2/p = 100$, and to contain a 
purely toroidal magnetic field. The energy flux $\dot{E}=10^{51}$ erg
s$^{-1}$ is kept constant during the entire simulation. For
simplicity, we neglect latitudinal variations in the wind and assume
that it is isotropic.  Self-consistency requires that the termination
shock produced by the interaction of the magnetar wind with the
surrounding star and MWN lie at $\simgt r_{min}$.  If not, the
assumption of a super-magnetosonic wind would not be valid.  We
discuss this constraint more in the next section.

We use a $35 M_\odot$ stellar progenitor from \citet{woo02}.  The wind
is injected inside a cavity of radius $\sim 10^9$ cm, which is roughly
the size of the collapsing region in the first second after core
bounce (our initial time).  In order to simulate the effect of a SN
shock propagating inside the progenitor, the region between $10^9$ and
$2\, 10^9$ cm is given an outward velocity corresponding to a total
kinetic energy $\sim 2\, 10^{51}$ erg, similar to that used in the
one-dimensional explosion calculations of \citet{ww95}.  The surface
of the progenitor star is located at $2.5\, 10^{10}$ cm. We assume
that the density outside the star falls off as $r^{-2}$; for the range
of radii we simulate, the results are independent of the outer density
profile.

\vspace{-0.7cm}

\section{Results}
\label{sec:res}

\begin{figure*}
\resizebox{\hsize}{!}{\includegraphics{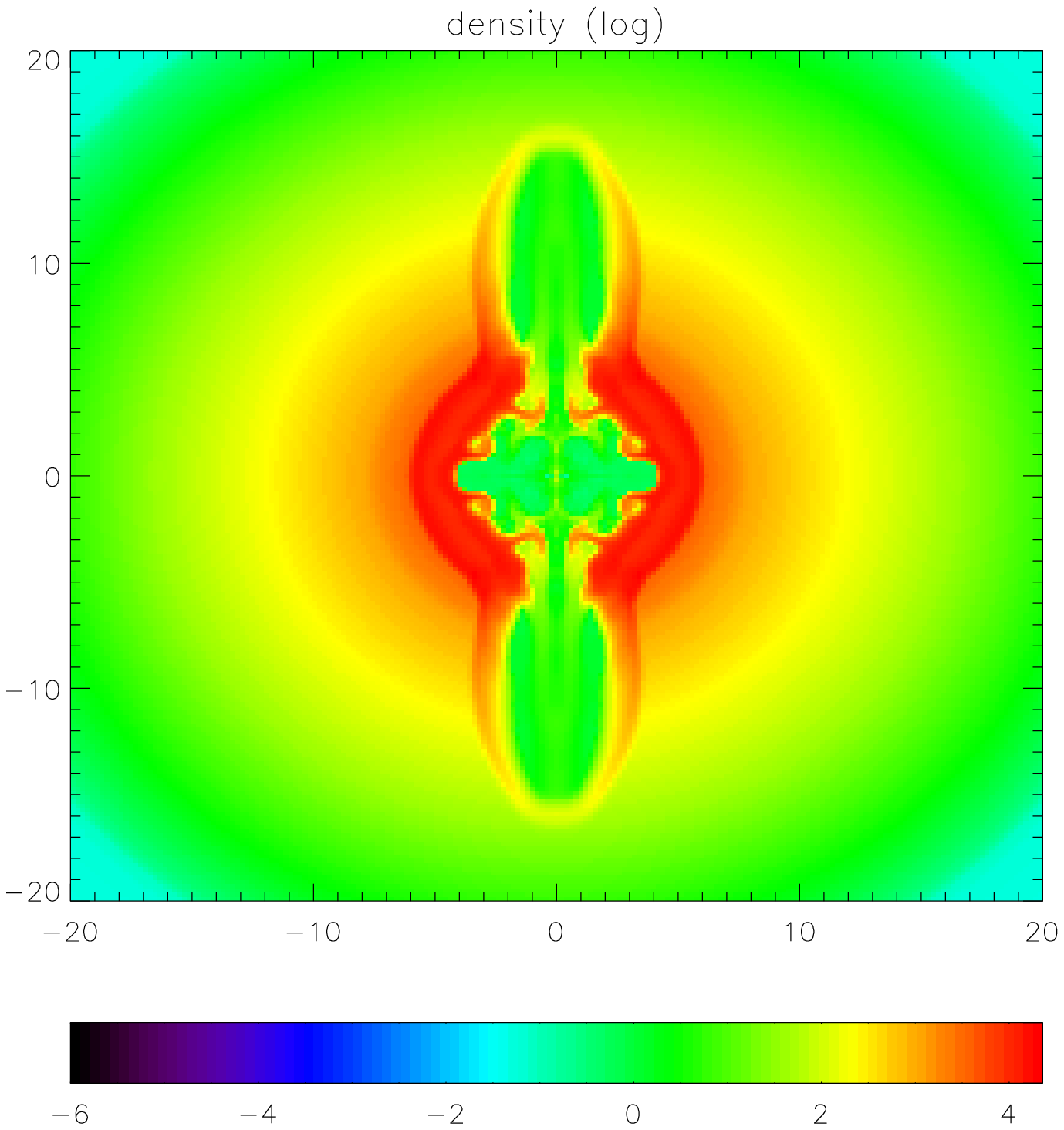}\includegraphics{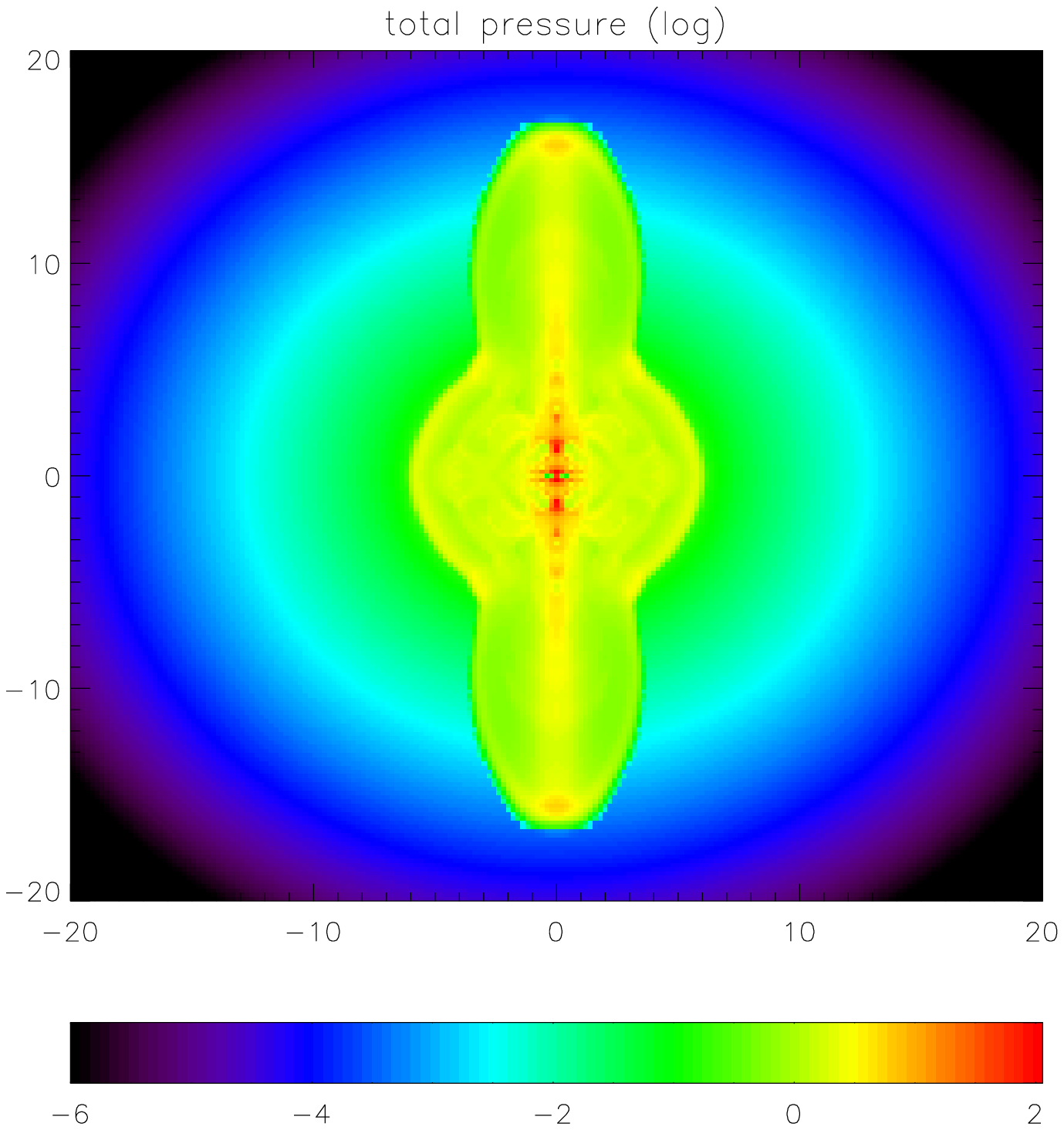}\includegraphics{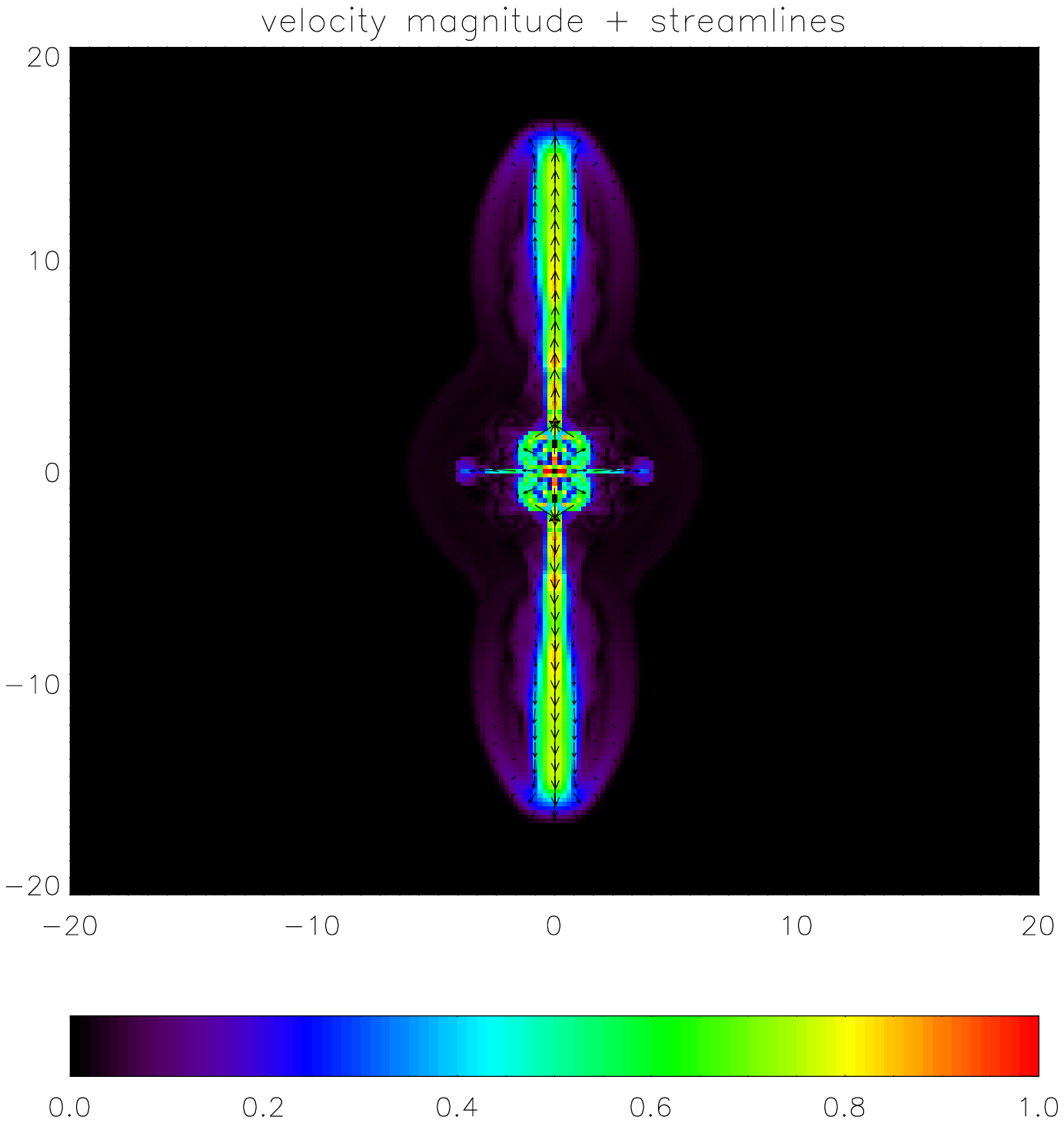}}
\resizebox{\hsize}{!}{\includegraphics{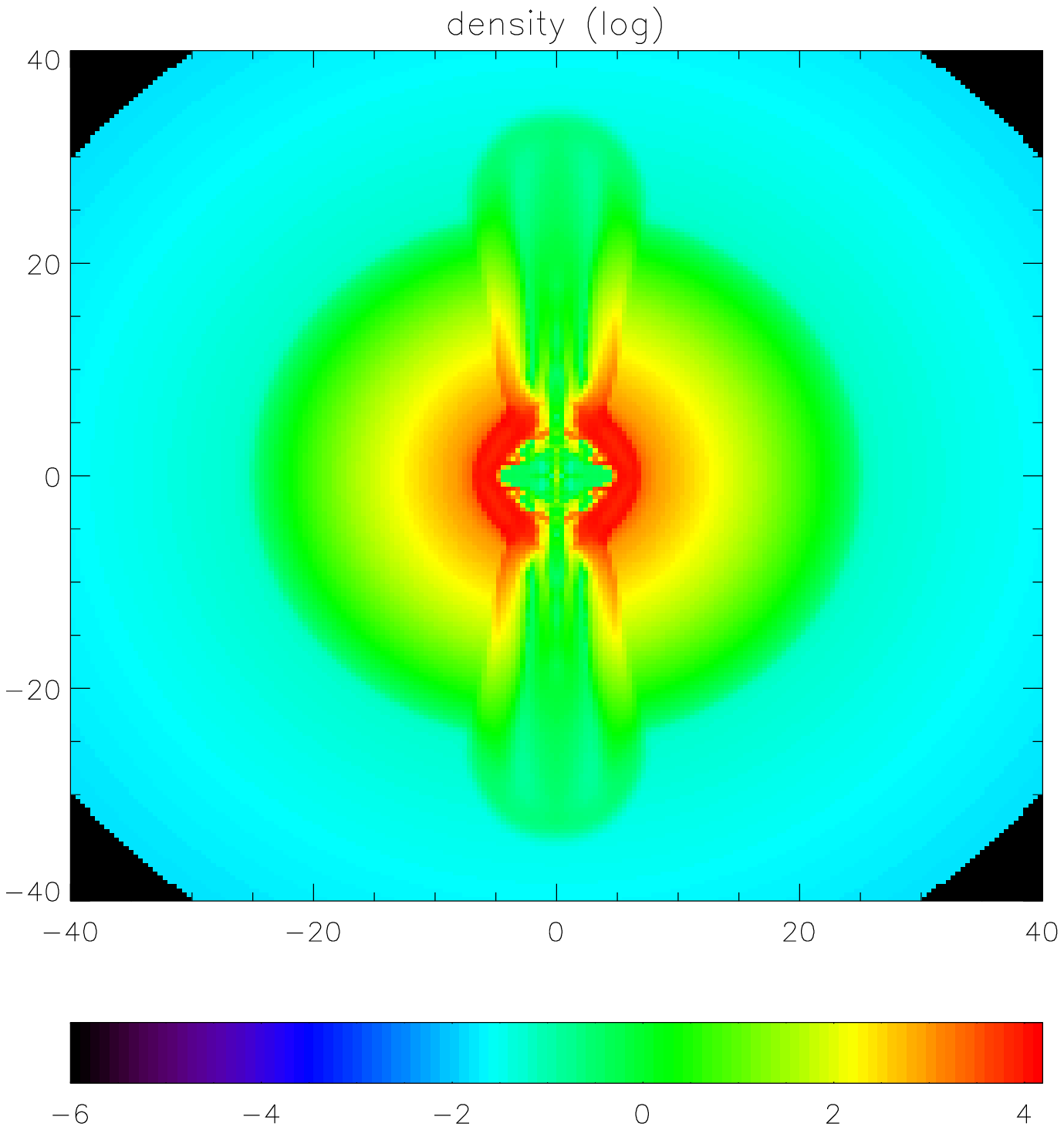}\includegraphics{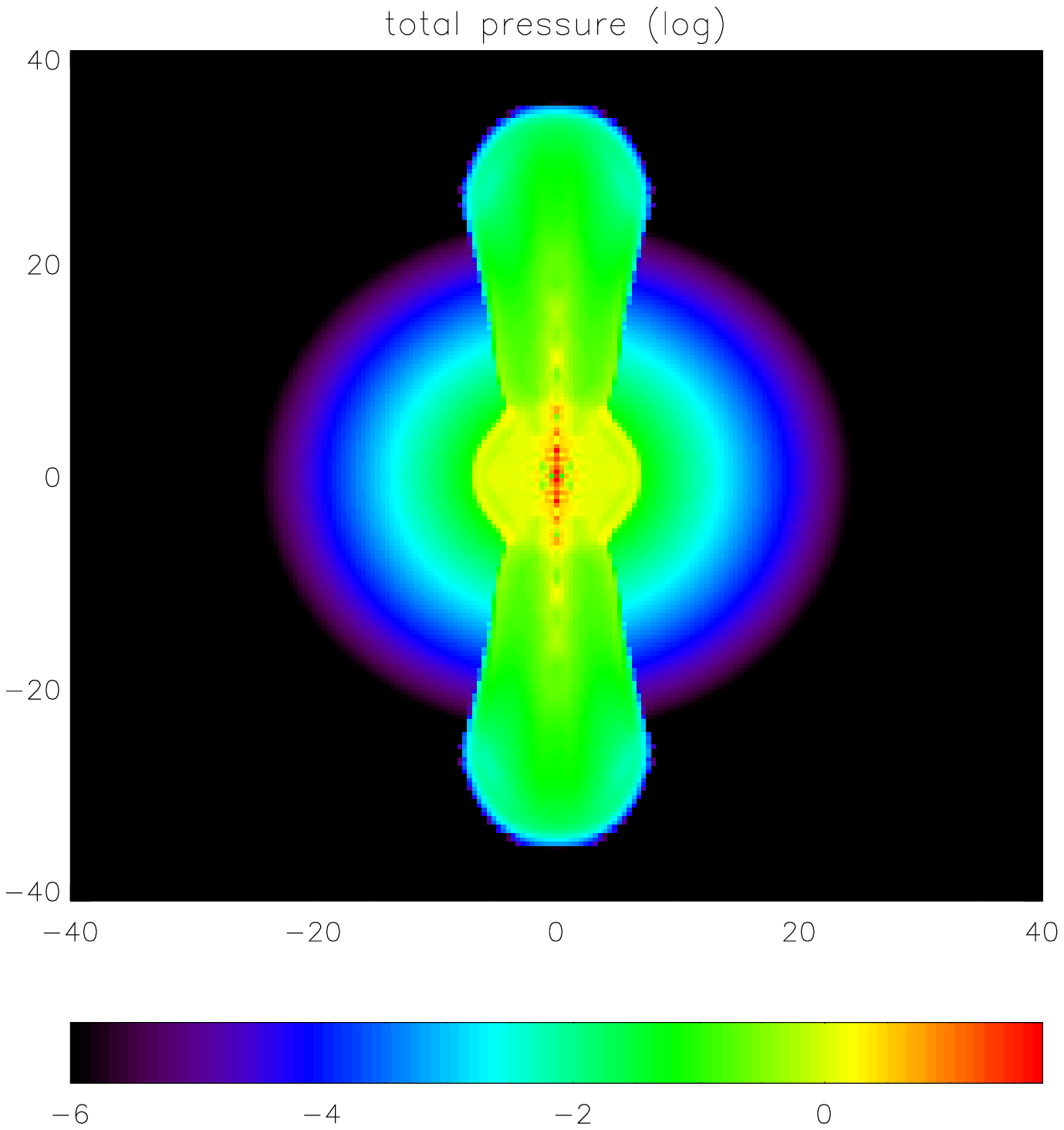}\includegraphics{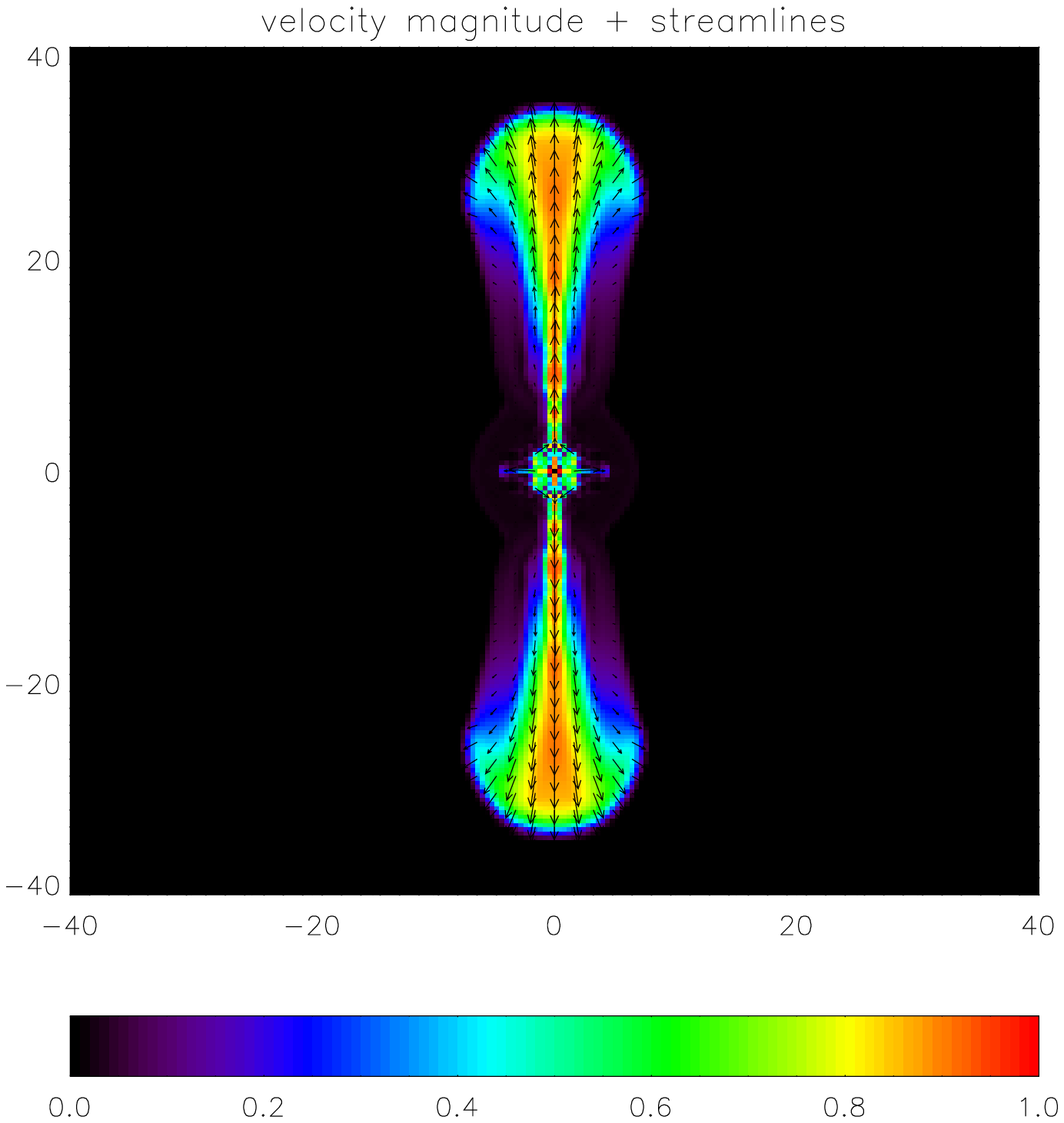}}
\resizebox{\hsize}{!}{\includegraphics{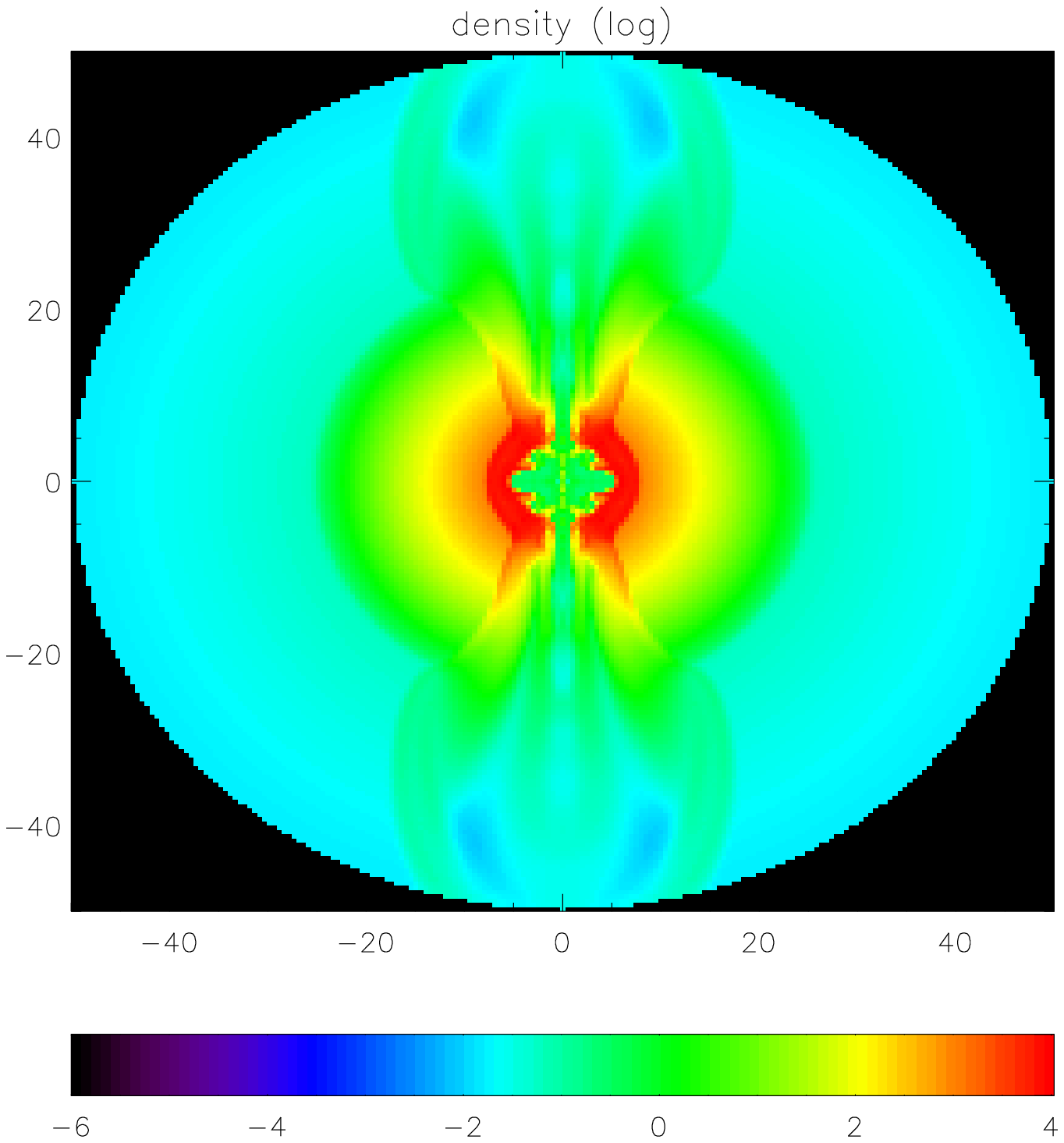}\includegraphics{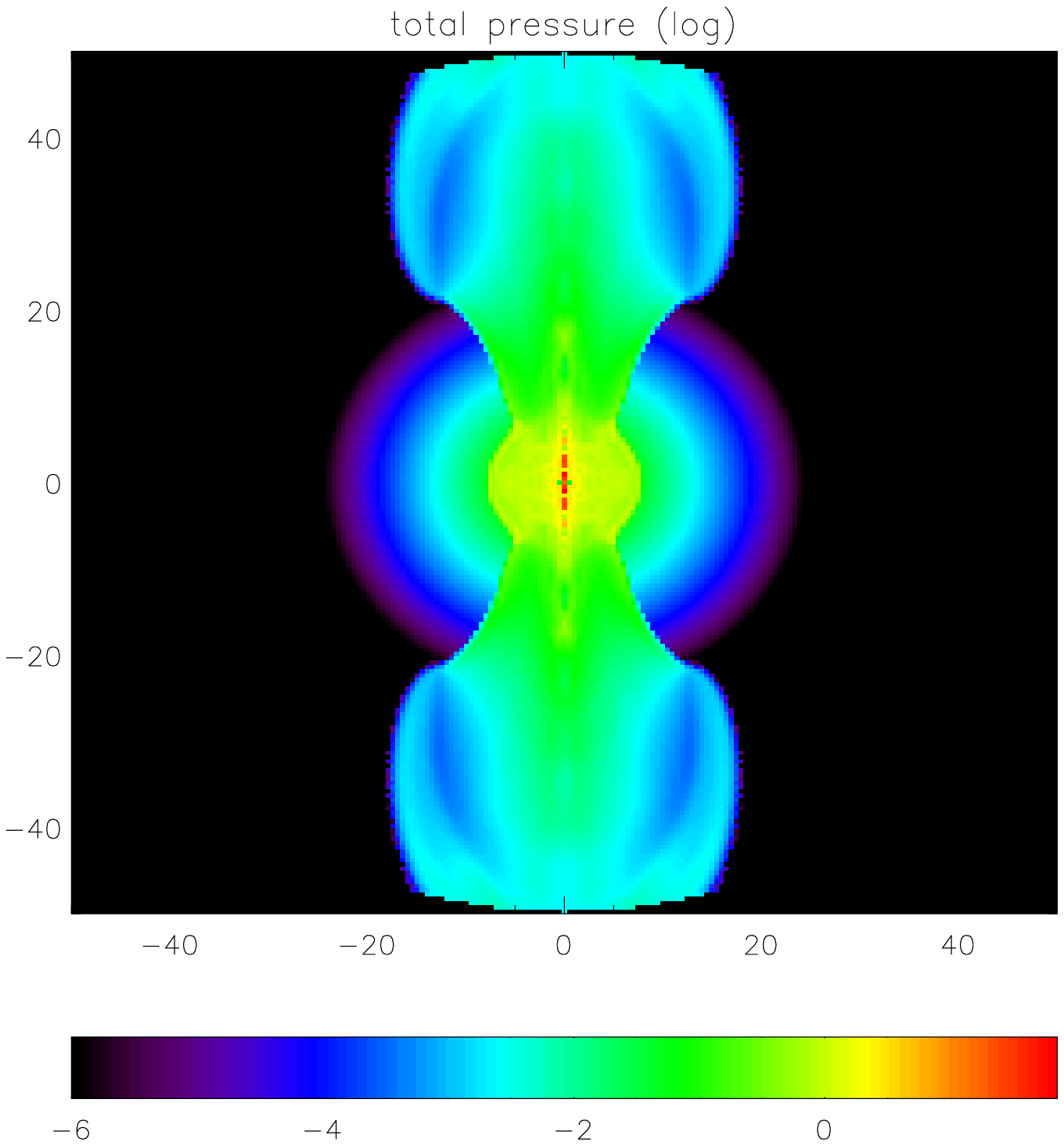}\includegraphics{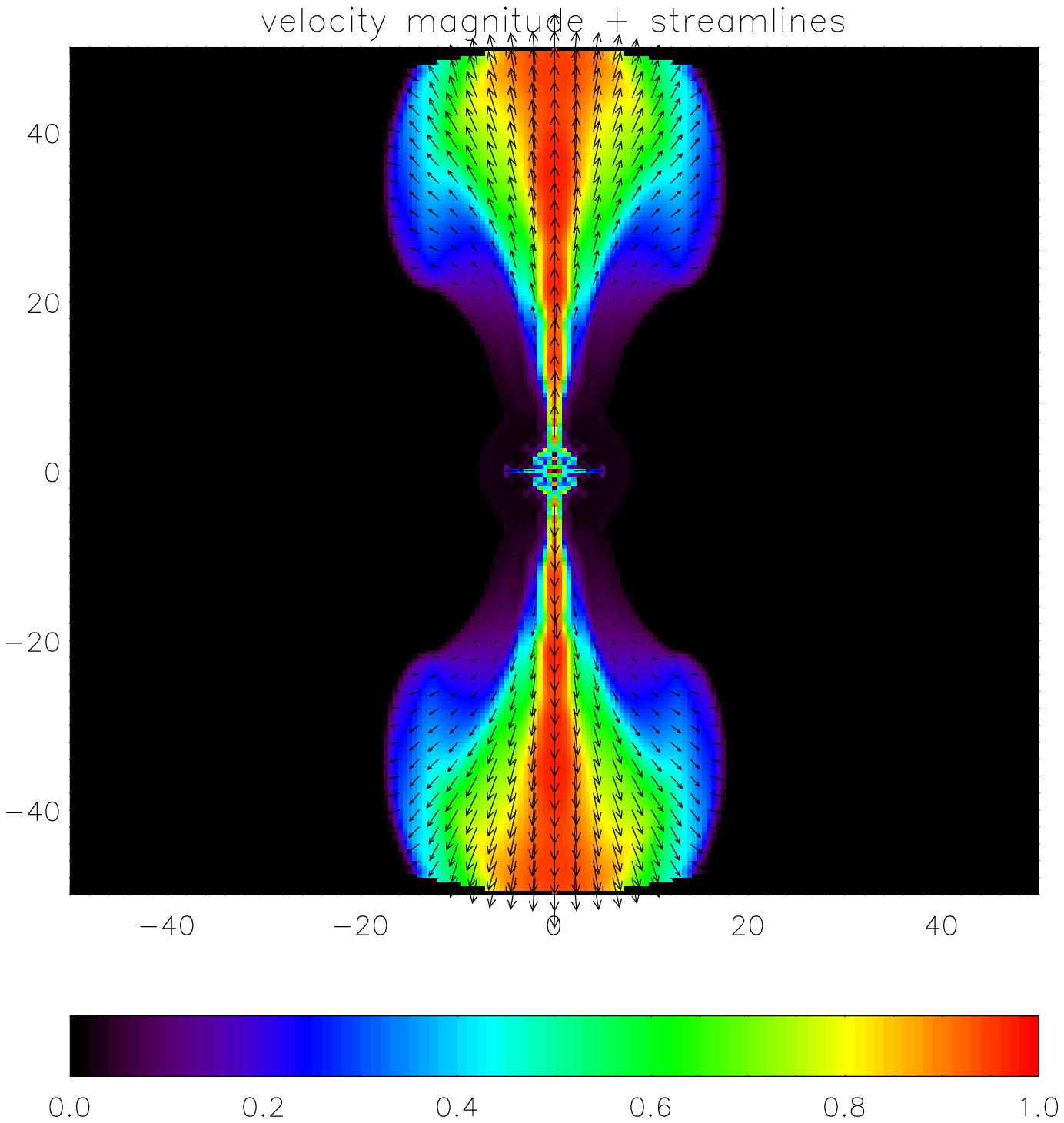}}
\caption{Evolution of a magnetized bubble inflated by a magnetar wind
with $\dot E = 10^{51} {\rm ergs \, s^{-1}}$, $\gamma_w = 10$, and
$\sigma = 0.1$, inside a $35 M_\odot$ progenitor star. From left to
right: density (g cm$^{-3}$), pressure (g cm$^{-3}\, c^2$) and
velocity (in units of $c$). From top to bottom: snapshots at $4,5,$ \&
$6$ seconds after core bounce. Distances are in 10$^9$ cm; the radius
of the progenitor star is $2.5 \, 10^{10}$ cm.  By t = 5 s (middle
panel) the jet has escaped the progenitor star.}
\label{fig:grb}
\end{figure*}

Figure \ref{fig:grb} shows the density, pressure, flow velocity and
flow streamlines for our simulations with $\gamma_w = 10$ at 3
different times, 4, 5, \& 6 seconds after core bounce.
 At all times there is an axial high velocity jet
subtending a few degrees surrounded by a somewhat less relativistic
cocoon subtending  $\sim 10^\circ$.  The jet carries
 a significant fraction of the
spindown energy of the magnetar.  Recall that the energy input in the
simulation is spherically symmetric, as is the progenitor star.  The
aspherical evolution of the system is generated self-consistently by
the dynamics of the MWN created inside the star.  Note that the
boundary between the MWN and the host star is particularly easy to
identify via the jump in the pressure.

\begin{figure}
\resizebox{\hsize}{!}{\includegraphics[bb = 0 0 450 430]{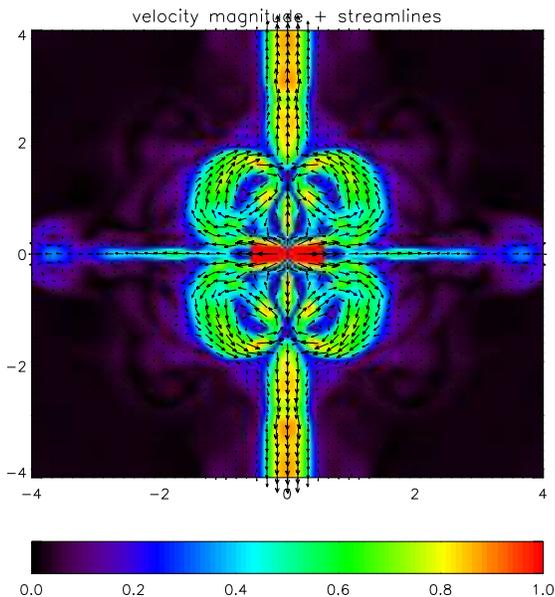}}
\caption{Inner velocity structure (in units of $c$) at 4 sec after core bounce. 
Distances are in 10$^9$ cm.  The termination shock is closer to
the source along the pole because the pressure in the bubble is larger
at the pole than at the equator. The flow pattern beyond the
termination shock shows the formation of an axial jet/plume due to
magnetic hoop stresses.  Because the magnetic field goes to 0 at the
equator by symmetry, there is a small region the equator where
material can escape in an equatorial channel.  This has no 
effect on the evolution of the system.}
\label{fig:inner}
\end{figure}

To help explain the physical origin of the jet in Figure
\ref{fig:grb}, Figure \ref{fig:inner} zooms in and shows the flow
velocity and streamlines in the inner region ($\simlt 4\,10^9$ cm) at
4 seconds.  The high velocity $v \approx c$ region at small radii in
Figure \ref{fig:inner} (indicated in red) is the freely expanding wind
with $\gamma_w = 10$. The wind goes through a termination shock at 
$\sim 10^8$ cm due to
the high bounding pressure of the MWN.  Even though
the energy flux in the freely expanding wind is isotropic, the
termination shock itself is highly asymmetric, with the radius of the
termination shock being significantly smaller along the pole.
Physically, this is because the strong toroidal field in the MWN
creates an anisotropic pressure distribution with the pressure at the
pole significantly larger than that at the equator \citep{beg92,b08}.
The higher polar pressure in turn causes the termination shock to be
at a smaller radius.  For the simulations in Figures \ref{fig:grb} \&
\ref{fig:inner}, with $\sigma = 0.1$ in the wind, we find that the
equatorial
termination shock always lies outside $\approx 10^8$ cm.  Thus our
injection of a super-magnetosonic wind is self-consistent.  By
contrast, in simulations with a larger value of $\sigma$ in the wind,
we found that the termination shock can move in to $\simlt r_{\min} =
10^7$ cm.  Because the termination shock would then lie within the
Alfv\'en surface, the spindown of the magnetar would be modified and
our assumption of super-magnetosonic injection would be invalid.  We
defer a more detailed study of this interesting possibility to future
work.

In the post termination shock region, Figure \ref{fig:inner} shows
that the flow undergoes a large-scale circulation with much of the
matter being diverted from the equatorial region to the pole where it
flows out along the jet, just as in analogous calculations for the
structure of pulsar wind nebulae \citep{kom04,ldz04}.  This is caused
by the toroidal magnetic field that builds up in the MWN, which is
nearly in equipartition near the location where the back-flow starts
\citep{ldz04}.  Although collimation by hoop stresses is ineffective
for the initially relativistic flow in the free wind, the
non-relativistic $v \sim 0.5 c$ flow in the post termination shock
region is effectively collimated by the magnetic field.

The MWN forms in about a sound crossing time $\sim 0.1-0.2$ sec.  It
takes $\sim 1$ second for the magnetar to fill the MWN with enough
energy for the magnetic stresses to become dynamically important,
 and about 1 more second for the jet/plume to emerge from 
the SN shock, still inside the star. The anisotropic
pressure distribution in the MWN, with a much larger pressure at the
pole than at the equator, and the significant energy and momentum flux
in the axial jet, act in concert to push out through the surrounding
star along the polar direction. At t = 4 sec, the MWN and jet are
still fully contained within the star (Fig. \ref{fig:grb}).  By t = 5
sec, however, they have reached the radius of the star at $\approx
2.5\,10^{10}$ cm and at t = 6 seconds, the material in the jet at
large radii has escaped the star and has accelerated back up to $v
\approx c$. Note that the high velocity core of the jet is surrounded
by a cocoon of less relativistic material that includes shocked
stellar wind material, as is generically expected to be the case
\citep{um06}.  The opening angle of the high $\gamma$ core of the jet
is a few degrees.  Because the jet is only marginally resolved in our
fiducial simulations, we ran a higher resolution simulation with twice
the number of grid points in the theta direction near the axis of the
jet; the results were nearly identical to those described here.

\begin{figure}
\resizebox{\hsize}{5cm}{\includegraphics{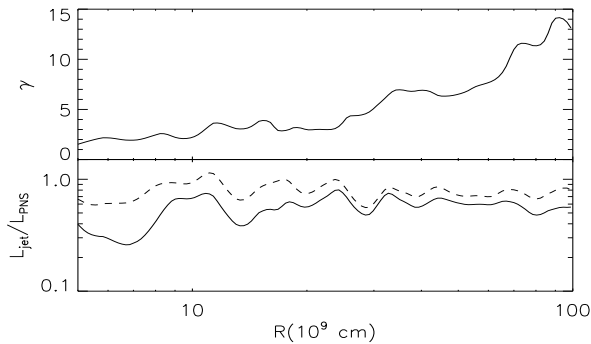}}
\caption{{\it Upper panel:} Lorentz factor on the axis of the jet, 9 sec after core bounce for $\gamma_w = 25$ and $r_{\max} = 10^{11}$ cm. $\gamma$
increases roughly linearly in radius. {\it Lower panel:} Ratio of the
radial energy flux in the high velocity core of the jet to the total
source power (solid line). Also shown is the ratio of the total radial
energy flux in the cocoon with respect to the source power (dashed
line).}
\label{fig:glf}
\end{figure}

Although the magnetic field is crucial for generating and collimating
the outflow seen in Figure \ref{fig:grb}, the field is not
energetically dominant; the ratio of the magnetic energy to the
thermal energy in the bubble is typically $\sim 0.1-0.2$, although it
can reach $\sim 1$ in the region where the hoop stress is most
effective (Fig. \ref{fig:inner}).  The enthalpy of the shocked gas in
the post-termination-shock region primarily determines the asymptotic
Lorentz factor in the jet.  In turn, the enthalpy of the shocked gas
is determined (through the termination shock's thermalization of the magnetar wind) by the wind Lorentz factor just upstream of the TS, $\gamma_w$.
Thus we find a roughly one-to-one relation between the Lorentz factor
of the jet core at large radii and $\gamma_w$: for $\gamma_w=10$ and
$r_{\rm max}=5\times10^{10}$ cm $\gamma$ reaches $\approx$ 5 in the
jet, whereas for $\gamma_w=25$ and $r_{\rm max}=10^{11}$ cm,
$\gamma\approx15$ at large radii.
The upper panel in Figure \ref{fig:glf} shows
$\gamma$ on the axis of the jet as a function of radius for this
simulation (at t = 9 sec), demonstrating the acceleration from the
marginally relativistic plasma at small radii near the termination
shock to the highly relativistic flow with $\gamma \approx \gamma_w$
at large radii.  The acceleration is roughly linear in radius,
consistent with that expected in a ``fireball'' with a constant
opening angle.

The lower panel in Figure \ref{fig:glf} shows the ratio of the power
in the high velocity core of the jet (solid) and in the wider cocoon
(dashed) relative to the power supplied by the central source $\dot
E$; as in the upper panel, these calculations are for $\gamma_w = 25$
at t = 9 sec when the outflow has escaped the star.  Figure
\ref{fig:glf} shows that at late times nearly all of the energy
supplied by the central magnetar escapes to large radii and that a
significant fraction of the energy is carried by the high velocity
core of the jet.

The upper and lower panels of Figure \ref{fig:glf} demonstrate radial
fluctuations in the Lorentz factor and $\dot E$, respectively, despite
the fact that the central source is assumed to have an $\dot E$ and
$\gamma_w$ that are constant in time.  These fluctuations arise
largely from shear instability near the base of the jet, and changes
in the circulation of matter in the MWN (the region seen in
Fig. \ref{fig:inner}) -- this accounts for the characteristic length
scale of $\sim 10^9$ cm in Figure \ref{fig:glf}.  The corresponding
variability timescale is $\sim 0.01-0.1$ sec.

\vspace{-0.75cm}

\section{Discussion \& Implications}
\label{sec:con}

Our simulations demonstrate that a wind with properties typical of 
proto-magnetar winds
inside a massive star can be collimated into an axial jet, even in the
conservative case in which the host star and  wind are
spherically symmetric.  Physically, this occurs because of the strong
toroidal magnetic field in the bubble of relativistic plasma and
magnetic field that is at first inertially confined by the progenitor
star and SN shock.  The magnetic field creates an asymmetric pressure
distribution in the bubble, with the pressure much larger at the pole
than the equator.  In addition, the toroidal field collimates the
moderately relativistic flow behind the termination shock into an
axial jet.

Although we have not been able to carry out simulations with Lorentz
factors of $\sim 100-10^3$ (largely for numerical reasons), as is
required to explain GRBs, extrapolating our results suggests that an
isotropic or equatorial flow with $\gamma \sim 100-10^3$ would
produce a jet at large radii with a comparable Lorentz factor.  This
collimated outflow would thus have physical properties similar to
those required to produce GRBs.  As noted in \S 1, calculations of the
mass-loss rate from newly formed magnetars find that magnetar winds
naturally have $\sigma_{LC} \simgt 100-10^3$ roughly 10-100 seconds after
core bounce (\citealt{thom04,met07}).

Figure \ref{fig:glf} shows that the jet escaping the star at large
distances has fluctuations in $\gamma$ and $\dot E$ on $\sim 0.01-0.1$
second timescales.  It is natural to
speculate that this variability in the source region could manifest
itself as variability in the gamma-ray emission in GRBs (via, e.g.,
internal shocks).  This is in addition to any intrinsic variability in
the spindown of the central magnetar.  It is also worth noting that
the magnetic field in the MWN will be advected out to large radii
where the GRB emission occurs \citep{lb03};
 the resulting field is likely strong
enough to account for the observed gamma-ray emission via synchrotron
radiation, without the need for shock generated magnetic fields.

This {\it Letter} has focused on one part of the parameter space of
proto-magnetar and host star interactions, namely a magnetar wind with
$\sigma \sim 0.1$ spinning down into a cavity evacuated by a
successful core-collapse SN.  We find that in simulations with a more
strongly magnetized wind ($\sigma \simgt 0.1$), the termination shock
moves inside the Alfv\'enic and fast magnetosonic surfaces (\S
\ref{sec:res}).  The spindown of the central magnetar will thus be
modified.  It is clearly of considerable interest to understand the
coupled dynamics of the magnetar wind and the surrounding nebula in
this limit.  In addition, in our current simulations, nearly all of
the spindown power of the central magnetar is channeled through the
polar jet (Fig. \ref{fig:glf}).  Thus the late-time spindown of the
magnetar can likely generate a GRB, but it will not energize the
surrounding SN.  It is thus not clear if the current model produces
{\it both} a hyper-energetic SN and a GRB, as is observed
(\citealt{woo06}).  One possibility is that the initial explosion
itself is highly energetic because rotational energy contributes to
the explosion (e.g., \citealt{thom05}).  Alternatively, we speculated
in \citet{b08} that because the initial magnetar wind is likely to be
significantly mass loaded and non-relativistic, it may contribute to
energizing the SN shock rather than to producing a GRB.  This
possibility will be investigated in future work with simulations that
properly take into account the evolution of $\dot E$ and $\sigma$ with
time in magnetar winds.  Finally, it is clearly important to understand the
stability of the strong toroidal field that is crucial to the dynamics
of the MWN and the axial jet. This can be studied both analytically
and with three dimensional simulations.

Near the completion of this work, \cite{k08} submitted a paper on
magnetar spindown inside a star whose results are complementary to
ours. They calculated the dynamics of an outflow from a magnetar 
interacting with an
infalling stellar progenitor for 200 ms after core-collapse, finding a
collimated non-relativistic jet-like outflow.  By contrast, we inject
a super-magnetosonic wind with properties derived from separate
spindown calculations, and are thus able to simulate the evolution for
a much longer period of time $\sim 10$ sec and to much higher Lorentz
factors.  Although a number of the details differ, the broad
conclusions from these two works are similar: magnetar formation can
produce a jet with properties similar to those required to produce
GRBs.

\vspace{-0.75cm}

\section*{Acknowledgments}
N.B. was supported by NASA through Hubble Fellowship grant
HST-HF-01193.01-A, awarded by the Space Telescope Science Institute,
which is operated by the Association of Universities for Research in
Astronomy, Inc., for NASA, under contract NAS 5-26555.  EQ and BDM
were supported by the David and Lucile Packard Foundation and a NASA
GSRP Fellowship to BDM. JA's research has
been supported by  NSF grant AST-0507813, NASA grant
NNG06GI08G, and DOE grant DE-FC02-06ER41453, all at UC Berkeley; by
the Department of Energy contract to the Stanford Linear Accelerator
Center no. DE-AC3-76SF00515; and by the taxpayers of California.
We thank A.~Heger for making massive stellar progenitor models 
available online.

\vspace{-0.75cm}


\label{lastpage}


\begin{thebibliography}{99}

\bibitem[\protect\citeauthoryear{Begelman \& Li}{1992}]{beg92}
Begelman, M.~C., \& Li, Z.\ 1992, \apj, 397, 187

\bibitem[\protect\citeauthoryear{Bucciantini et al.}{2006}]{me06}
 Bucciantini, N., Thompson, T.~A., Arons, J., Quataert, E., \& Del
 Zanna, L.\ 2006, \mnras, 368, 1717

\bibitem[\protect\citeauthoryear{Bucciantini et al.}{2007}]{b08}
Bucciantini, N., Quataert, E., Arons, J., Metzger, B.~D., \&
Thompson, T.~A.\ 2007, ArXiv e-prints, 705, arXiv:0705.1742 

\bibitem[\protect\citeauthoryear{Del Zanna \& Bucciantini}{2002}]{ldz02}
 Del Zanna, L., \& Bucciantini, N.\ 2002, \aap, 390, 1177 

\bibitem[\protect\citeauthoryear{Del Zanna et al.}{2003}]{ldz03}
 Del Zanna, L., Bucciantini, N., \& Londrilo, P.\ 2003, \aap, 400, 397 

\bibitem[\protect\citeauthoryear{Del Zanna et al.}{2004}]{ldz04}
 Del Zanna, L., Amato, E., \& Bucciantini, N.\ 2004, \aap, 421, 1063 

\bibitem[\protect\citeauthoryear{Komissarov \& Lyubarsky}{2004}]{kom04}
 Komissarov, S.~S., \& Lyubarsky, Y.~E.\ 2004, \mnras, 349, 779 

\bibitem[\protect\citeauthoryear{Komissarov \& Barkov}{2007}]{k08}
Komissarov, S.~S., \& Barkov, M.~V.\ 2007, ArXiv e-prints, 707,
arXiv:0707.0264

\bibitem[K{\"o}nigl \& Granot(2002)]{kg02}
 K{\"o}nigl, A., \& Granot, J.\ 2002, \apj, 574, 134  

\bibitem[\protect\citeauthoryear{Lyubarsky \& Eichler}{2001}]{le01} 
Lyubarsky, Y.~E., \& Eichler, D.\ 2001, ApJ, 562, 494 

\bibitem[Lyutikov \& Blandford(2003)]{lb03}
 Lyutikov, M., \& Blandford, R.\ 2003, ArXiv e-prints, arXiv:astro-ph/0312347
 
\bibitem[\protect\citeauthoryear{MacFadyen \& Woosley}{1999}]{mcf99}
 MacFadyen, A.~I., \& Woosley, S.~E.\ 1999, \apj, 524, 262 

\bibitem[\protect\citeauthoryear{Matzner}{2003}]{mat03}
 Matzner, C.~D.\ 2003, \mnras, 345, 575 

\bibitem[\protect\citeauthoryear{Metzger et al.}{2007}]{met07}
 Metzger, B.~D., Thompson, T.~A., \& Quataert, E.\ 2007, ApJ, 659, 561

\bibitem[\protect\citeauthoryear{Rhoads}{1999}]{rho99}
 Rhoads, J.~E.\ 1999, \apj, 525, 737

\bibitem[\protect\citeauthoryear{Sherwin \& Lynden-Bell}{2007}]{she07}
 Sherwin, B.~D., \& Lynden-Bell, D.\ 2007, \mnras, 378, 409

\bibitem[\protect\citeauthoryear{Thompson}{1994}]{thom94} 
Thompson, C.\ 1994, MNRAS, 270, 480 

\bibitem[\protect\citeauthoryear{Thompson et al.}{2004}]{thom04}
Thompson, T.~A., Chang, P., Quataert, E.\ 2004, ApJ, 611, 380

\bibitem[\protect\citeauthoryear{Thompson et al.}{2005}]{thom05}
Thompson, T.~A., Quataert, E., \& Burrows, A.\ 2005, \apj, 620, 861

\bibitem[\protect\citeauthoryear{Usov}{1992}]{usov92} 
Usov, V.~V.\ 1992, Nature, 357, 472 

\bibitem[\protect\citeauthoryear{Uzdensky \& MacFadyen}{2006}]{um06}
 Uzdensky, D.~A., \& MacFadyen, A.~I.\ 2006, \apj, 647, 1192

\bibitem[\protect\citeauthoryear{Uzdensky \& MacFadyen}{2007}]{um07} 
Uzdensky, D. A. \& MacFadyen, A. I., 2007, ApJ accepted (astro-ph/0609047)

\bibitem[\protect\citeauthoryear{Wheeler et al.}{2000}]{wheeler00} 
Wheeler, J.~C., Yi, I., H{\"o}flich, P., \& Wang, L.\ 2000, ApJ, 537, 810 

\bibitem[\protect\citeauthoryear{Woosley \& Bloom}{2006}]{woo06}
 Woosley, S.~E., \& Bloom, J.~S.\ 2006, ARA\&A, 44, 507

\bibitem[\protect\citeauthoryear{Woosley et al.}{2002}]{woo02}
 Woosley, S.~E., Heger, A., \& Weaver, T.~A.\ 2002, Reviews of Modern
 Physics, 74, 1015

\bibitem[\protect\citeauthoryear{Woosley \& Weaver}{1995}]{ww95}
Woosley, S.~E., \& Weaver, T.~A.\ 1995, ApJS, 101, 181


\end{thebibliography}
\end{document}